# Flaw-tolerant recycling of waste tires into high-performance rubber

Chenghai Li,[1,4] Yu Zhou,[1,4] Zheqi Chen,[1,4] Jun Ho Min[1], Matthew Wei Ming Tan,[1,2] Masako Yoshihara Yang,[1] Yakov Kutsovsky[3] and Zhigang Suo[*1]

[1] John A. Paulson School of Engineering and Applied Sciences, Harvard University, Cambridge, MA 02138, USA
[2] School of Materials Science and Engineering, Nanyang Technological University, 50 Nanyang Avenue, Singapore 639798, Singapore
[3] Expert-in-Residence, Office of Technology Development, Harvard University, Cambridge, MA 02138, USA
[4] These authors contributed equally to this work.
[*] To whom correspondence should be addressed. Email: suo@seas.harvard.edu (Z.S.)

## Abstract
Annually about one billion tires reach the end of life. Rubber from waste tires can be ground into particles and incorporated into a matrix. In practice, the resulting composites are commonly not *recycled* into new tires, but *downcycled* into low-value products such as sport surfaces, indoor flooring, and rubberized asphalt. Recycling waste tires into new tires remains a challenge. In particular, such composites typically have low strength, especially when the particle content is high or the particles are large. A single waste tire particle may act as a flaw to initiate macroscopic fracture and knock down the strength of the entire composite. Here we propose the principle of flaw-tolerant recycling. Rather than requiring every particle to be reliable and small, the matrix is designed so that flaw-like particles do not initiate macroscopic fracture. We demonstrate a flaw-tolerant matrix consisting of two intertwined networks: a network of densely entangled and sparsely crosslinked long natural rubber chains, and a percolated network of carbon aggregates. When the flaw-tolerant matrix is incorporated with up to 80 wt% of commercially available waste particles, the resulting composites still maintain strength. The principle of flaw-tolerant recycling may aid the development of recycling waste tires into new tires.

## Keywords
Waste tire recycling; Flaw tolerance; Rubber composite; Strength; Fracture; Particle size; Particle content

## 1. Introduction

Over half of waste tires are burned or landfilled[1-3]. Turning waste tires into useful products would promote a circular economy[4-8]. A common route is to separate waste tires into rubber, steel and textile fibers, and grind the rubber into particles[7,9,10]. The waste tire particles are then incorporated into a matrix[7,9-11]. In practice, such composites are often not *recycled* into new tires, but *downcycled* into low-value products, such as sport surfaces, indoor flooring, and rubberized asphalt (Figure 1a). Efforts have long been made to recycle waste tires into new tires[1,12-15]. The composites made of waste tire particles, however, often have low mechanical performance, especially when the particles are large or their content is high[7,12,16,17].

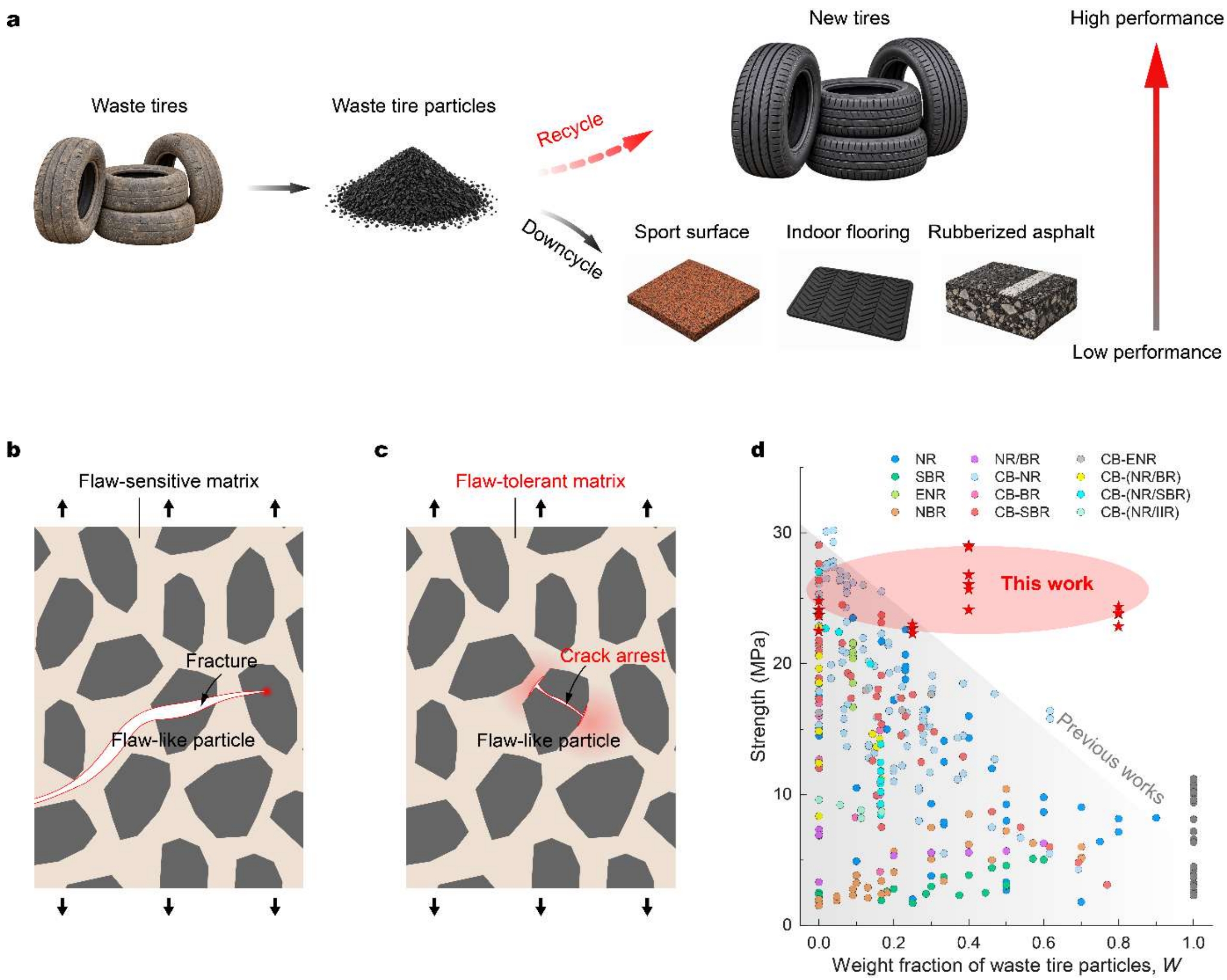

**Figure 1. Flaw-tolerant recycling of waste tires into high-performance rubber**. (**a**) Rubber from waste tires can be ground into particles and then incorporated into a matrix. The resulting composites, however, often have low mechanical performance and are commonly not used in new tires, but are used in low-value products such as sport surfaces, indoor flooring, and rubberized asphalt. (**b**) In a flaw-sensitive matrix, flaw-like particles initiate crack growth. (**c**) In a flaw-tolerant matrix, the same flaw-like particles do not initiate crack growth. (**d**) When $W = 0$, the data are for the matrix. When $W = 1$, the data are for waste tire rubber. When $0 < W < 1$, the data are for composites made of waste tire particles using different matrices. The matrices include natural rubber (NR), styrene-butadiene rubber (SBR), epoxidized natural rubber (ENR), nitrile butadiene rubber (NBR), natural rubber/butadiene rubber (NR/BR), carbon black filled natural rubber (CB-NR), carbon black filled butadiene rubber (CB-BR), carbon black filled styrene-butadiene

rubber (CB-SBR), carbon black filled epoxidized natural rubber (CB-ENR), carbon black filled natural rubber/butadiene rubber (CB-NR/BR), carbon black filled natural rubber/styrene-butadiene rubber (CB-NR/SBR), carbon black filled natural rubber/isobutylene-isoprene rubber (CB-NR/IIR). Data for previous works are compiled from the literature; the corresponding sources are listed in Table S1.

The reduction of mechanical performance originates from two characteristics of waste tire particles. First, during service, tires undergo oxidation and cyclic deformation, so that waste tire particles have degraded mechanical properties with large variability from particle to particle[18-21]. Second, fine grinding is costly, so that commercially available waste tire particles are relatively large with a broad size distribution[5,7,16,17]. To improve the mechanical properties of composites containing waste tire particles, extensive work has been devoted to enhancing particle-matrix adhesion[22-26]. However, even with strong particle-matrix adhesion, weak particles themselves may act as flaws. The characteristic flaw size scales with the particle size. A single flaw-like particle can initiate macroscopic fracture and knock down the strength of the entire composite (Figure 1b). Consequently, recycling waste tires into new tires requires not only strong particle-matrix adhesion, but also a flaw-tolerant matrix.

Here we propose the principle of flaw-tolerant recycling. Rather than requiring every particle to be reliable and small, we design a matrix to tolerate flaw-like particles, so they do not initiate the macroscopic fracture of the composite (Figure 1c). Specifically, a flaw-tolerant matrix can tolerate particles of moderate size and high weight fraction. We demonstrate composites that maintain strength even with ~80 wt% of commercially available waste tire particles (Figure 1d). By comparison, previously reported composites substantially lose strength when the weight fraction of particles exceeds ~20 wt%. In addition to strength, the composites demonstrated in this work also maintain stiffness, stretchability, and work of fracture (Figure S1). The flaw-tolerant principle broadens the allowable particle size while maintaining properties, thereby reducing the need to grind waste tire rubber into extremely fine particles.

# 2. Results

## 2.1 Flaw-tolerant matrix

As a proof of concept, here we demonstrate a flaw-tolerant matrix consisting of two intertwined networks: a percolated network of carbon aggregates, and a network of densely entangled and sparsely crosslinked long natural rubber chains (Figure 2a-c). The long natural rubber chains are preserved through a low-intensity wet-mixing process[27-29], in which natural rubber latex, carbon aggregates, and a thermal initiator are mixed with a solvent (Figures S2 and S3). As the solvent evaporates, the rubber chains entangle among themselves and intertwine with carbon aggregates. The dried compound is shaped and cured by hot pressing. During curing, the thermal initiators decompose into free radicals that create sparse crosslinks between long rubber chains. Consequently, the strands between neighboring crosslinks are long.

For comparison, we also prepare a matrix with short rubber chains as follows (Figure S4). The compound is prepared using the same processes of wet-mixing and drying, followed by high-intensity roll milling. The milled compound is then shaped and cured by hot pressing. Roll milling cuts the natural rubber chains short, and the resulting dangling ends of the short chains do not bear load[30,31]. Consequently, dense crosslinks are required to form a proper network, resulting in short strands between neighboring crosslinks[30,31] (Figure 2d).

Both matrices contain the same volume fraction of carbon aggregates, $F = 0.25$. The initiator fraction, $C$, is defined as the molar ratio of the thermal initiator to the repeat units per polyisoprene chain. A low initiator fraction, $C = 2\times10^{-3}$, is chosen for the matrix with long chains, and a high initiator fraction, $C = 10^{-2}$, is chosen for the matrix with short chains. The two matrices have similar stress-stretch curves and comparable tensile properties, including modulus, strength, and work of fracture (Figure 2e-h). Here, the work of fracture, $W_c$, is defined as the area under the stress-stretch curve up to rupture.

The distinction between the two matrices becomes pronounced in the presence of flaws. For a notched sample under monotonic loading, the initiation of crack growth occurs at $\lambda = 1.67$ in the matrix with long chains, but at only $\lambda = 1.35$ in the matrix with short chains (Figure 2i). The corresponding energy release rate defines the initiation toughness, $G_i$ (Figure 2j, and Figures S5 and S6). The matrix with long chains has a much higher $G_i$ than the matrix with short chains. Such an elevated initiation toughness is attributed to the synergy of long strands, strain-induced crystallization, as well as interfacial slip and desorption between carbon aggregates and rubber chains[29] (Figure S7).

The ratio of the initiation toughness to work of fracture, $G_i/W_c$, defines a material-specific length, called the fractocohesive length[32,33]. This characteristic length measures the flaw sensitivity of a material. The material is insensitive to flaws that are small compared with the fractocohesive length, but is sensitive to flaws that are large compared with the fractocohesive length (Figure 2k). The $W_c$ of the matrix with long chains is approximately 1.27 times that of the matrix with short chains, but the $G_i$ of the former is approximately 3.83 times that of the latter. Consequently, the matrix with long chains has a much larger fractocohesive length, $G_i/W_c \sim 270$ µm, compared with $G_i/W_c \sim 90$ µm for the matrix with short chains (Figure 2l). We therefore refer to the matrix with long chains as a flaw-tolerant matrix, and the matrix with short chains as a flaw-sensitive matrix. As we will show below, this enlarged fractocohesive length enables the incorporation of larger particles and higher weight fractions of particles while maintaining strength.

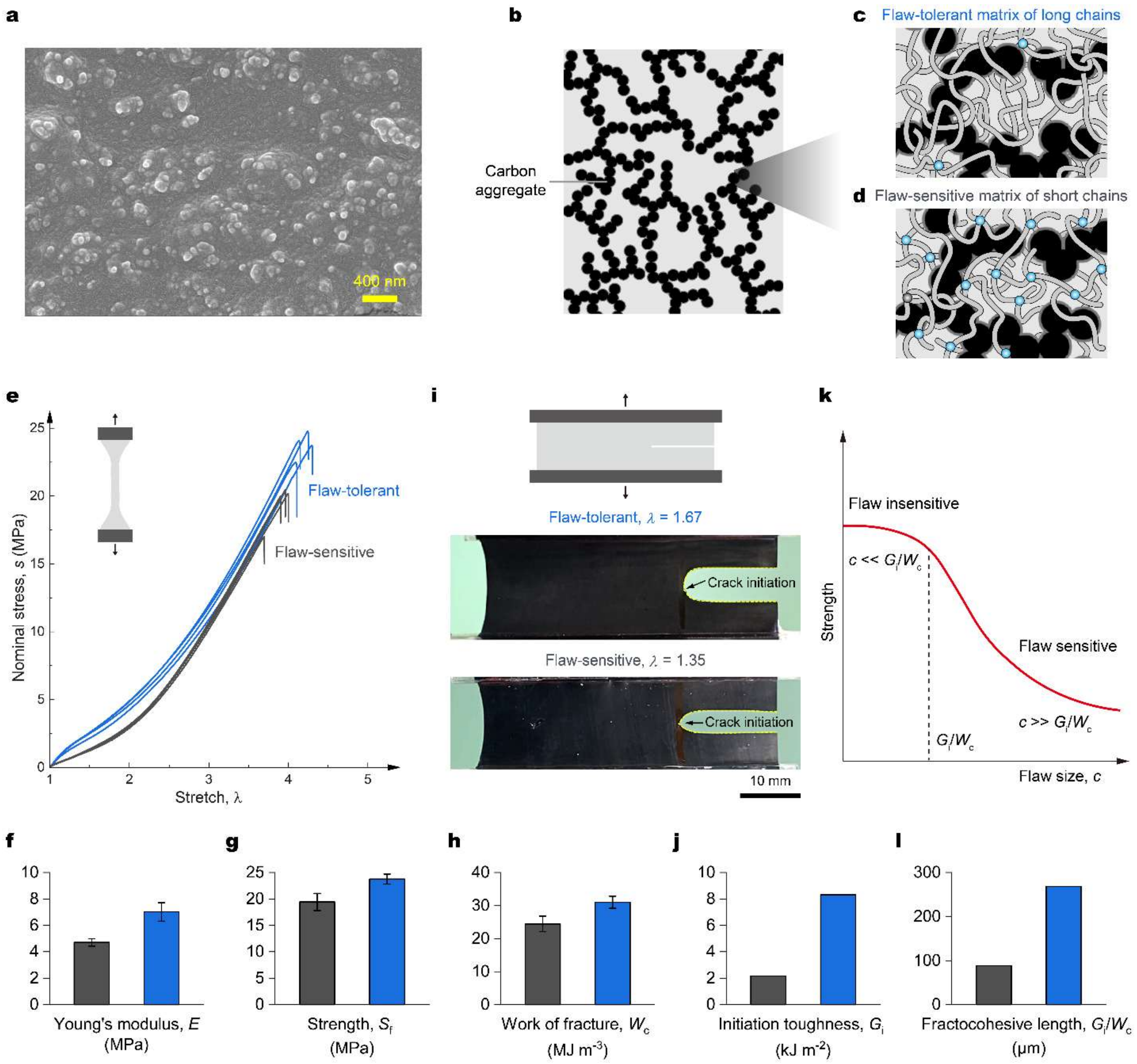

**Figure 2. Flaw-tolerant and flaw-sensitive matrices. (a)** Scanning electron microscopy (SEM) image of the cross section of the flaw-tolerant matrix. A flaw-tolerant matrix consists of **(b)** a percolated network of carbon aggregates, and **(c)** a network of densely entangled and sparsely crosslinked long natural rubber chains. **(d)** In a flaw-sensitive matrix, short rubber chains are densely crosslinked to form a proper network. **(e)** Stress-stretch curves. **(f)** Young's modulus, $E$. **(g)** Strength, $S_f$. **(h)** Work of fracture, $W_c$. **(i)** Initiation of crack growth in notched samples. **(j)** Initiation toughness, $G_i$. **(k)** Schematic illustrating the fractocohesive length, $G_i/W_c$. **(l)** Measured values of $G_i/W_c$. Both matrices have the same volume fraction of carbon aggregates, $F = 0.25$. The initiator fractions, $C$, for the flaw-tolerant and flaw-sensitive matrices $2\times10^{-3}$ and $10^{-2}$, respectively. Values in **(f)**, **(g)**, and **(h)** represent averages with standard deviations for $N = 4$ replicates.

## 2.2 A flaw-tolerant matrix recycles large waste tire particles while maintaining strength

We next demonstrate that a flaw-tolerant matrix enables the recycling of large waste tire particles while maintaining strength. Commercially available waste tire particles are separated into different size ranges, with the average size ranging from 74 μm to 750 μm (Figure S8). Then, waste tire particles, carbon aggregates, natural rubber latex, and a thermal initiator are mixed in a solvent (Figure S9). After the solvent evaporates, the compound is shaped and cured by hot pressing. During curing, rubber chains become mobile and diffuse into waste particles, forming interfacial entanglements. Meanwhile, the thermal initiators decompose into free radicals that crosslink fresh rubber chains and interlink fresh rubber chains with waste particles. In the cured composite, the matrix preserves the same two intertwined networks, and adheres to waste particles through interlinks and interfacial entanglements (Figure 3a,b). A representative SEM image confirms that the waste particles are uniformly distributed throughout the matrix (Figure 3c).

For comparison, we also prepare composites with a flaw-sensitive matrix (Figure S10). The natural rubber latex is first dried and then roll-milled to cut rubber chains short. Next, the milled rubber is dissolved in a solvent, after which a thermal initiator, carbon aggregates, and waste particles are added. After the solvent evaporates, the compound is shaped and cured by hot pressing.

We next examine the effect of particle size on the strength of the composites. For the composite with a flaw-sensitive matrix, the strength is maintained only for relatively small particles with an average size of ~74 μm (Figure 3d and Figure S11). The strength decreases markedly when the average particle size reaches ~180 μm, and decreases further for particles with an average size of ~400 μm. By contrast, for the composite with a flaw-tolerant matrix, the strength is maintained for average particle sizes up to ~180 μm (Figure 3e and Figure S12). The strength begins to decrease when the average particle size is increased to ~275 μm. For extremely large particles with the average size of ~750 μm, the strength remains around 12 MPa.

We summarize the dependence of strength on the average particle size for the two composites (Figure 3f). The composite with a flaw-tolerant matrix maintains high strength over a broader range of particle sizes than the composite with a flaw-sensitive matrix, demonstrating a greater tolerance to flaw-like particles.

We interpret the experimental results as follows. Waste tire particles have degraded mechanical properties and large variability from particle to particle[18-21]. Even if a single particle debonds from the matrix or fractures internally, the particle acts like a flaw. The characteristic flaw size scales with the particle size. When the particle size is large compared to the fractocohesive length of the matrix, the flaw-like particle initiates crack growth and reduces the strength of the composite. By contrast, when the particle size is small compared to the fractocohesive length, the flaw-like particle cannot initiate crack growth, and the strength of the composite is maintained. Consequently, the large fractocohesive length of the flaw-tolerant matrix broadens the range of particle sizes that can be incorporated while maintaining strength.

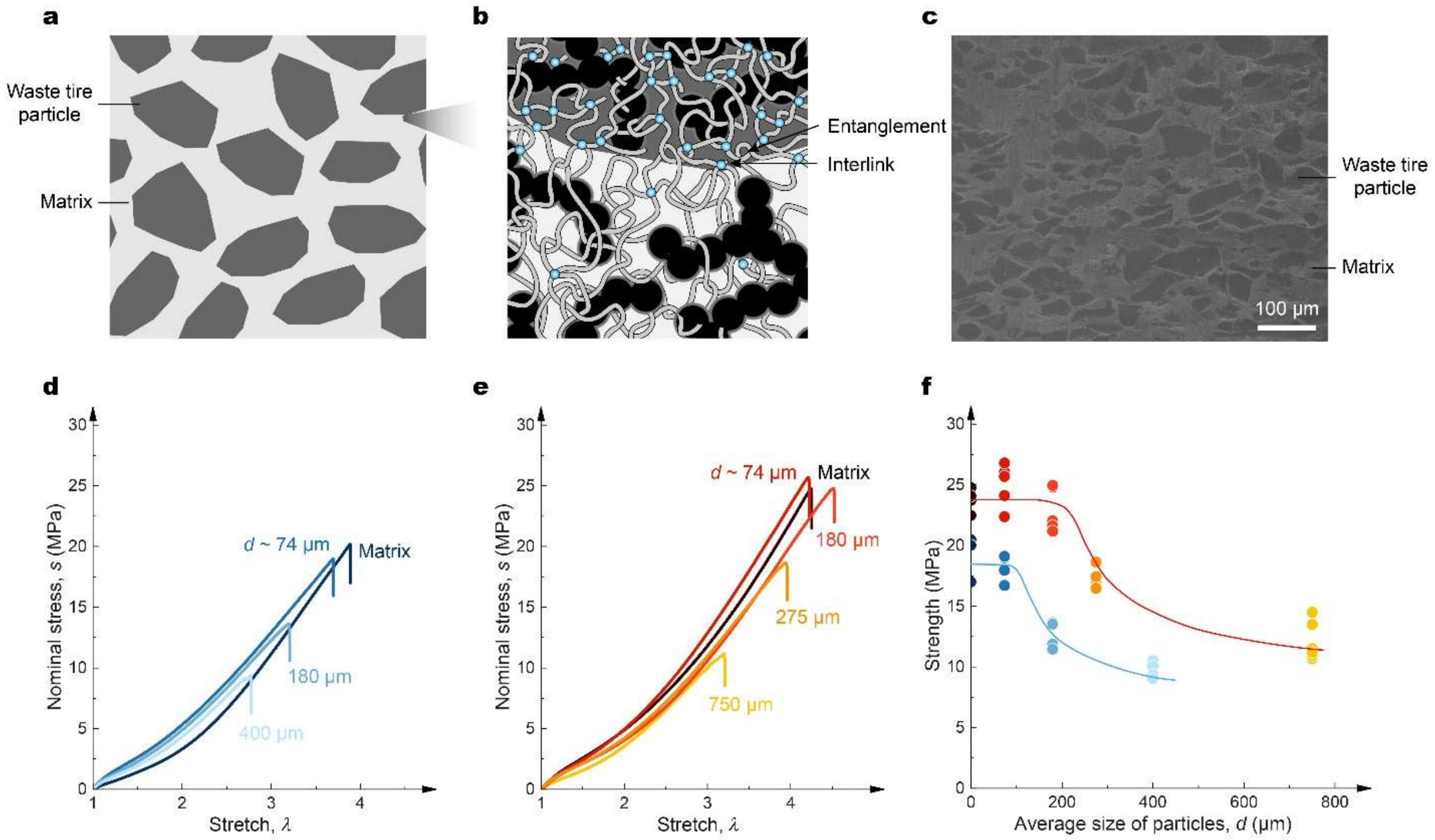

**Figure 3. A flaw-tolerant matrix enables the recycling of large waste tire particles while maintaining strength. (a)** Schematic of waste tire particles distributed in the matrix. **(b)** The matrix adheres to particles through interlinks and entanglements. **(c)** SEM image of the cross section of the composite. Stress-stretch curves of composites containing particles of different sizes with **(d)** flaw-sensitive and **(e)** flaw-tolerant matrices. **(f)** Strength as a function of the particle size. Both matrices have the same volume fraction of carbon aggregates, $F = 0.25$. The initiator fractions for the flaw-tolerant and flaw-sensitive matrices are $C = 2\times10^{-3}$ and $C = 10^{-2}$, respectively. Both composites have the same weight fraction of particles, $W = 0.4$.

## 2.3 A flaw-tolerant matrix recycles a high fraction of waste tire particles while maintaining strength

We next demonstrate that a flaw-tolerant matrix enables the recycling of a high fraction of waste tire particles while maintaining strength, provided that the particle size is smaller than the fractocohesive length. We prepare composites with different fractions of particles and examine their cross sections by SEM (Figure 4a). The average particle size is 74 μm. For the matrix, $W = 0$, the cross section is homogeneously gray. As the particle fraction increases from $W = 0.25$ to $W = 0.4$, the particles (dark regions) occupy an increasing fraction, whereas the matrix (gray regions) occupies a decreasing fraction. At an extremely high particle fraction, $W = 0.8$, the particles are densely packed, and the matrix between neighboring particles becomes very thin, such that the boundaries of individual particles become less distinct.

The particle fraction of $W = 0.8$ corresponds to a volume fraction of approximately 80%, far exceeding the packing limit of rigid monodisperse spheres[34-36]. Such a high fraction is realized because the particles are irregular in shape, broadly distributed in size, and deformable. During hot pressing, neighboring particles deform and conform to one another, reducing the interparticle void space and enabling high packing fractions. Even at such a high fraction of particles, neighboring particles remain separated by thin layers of the matrix.

Despite the high fraction of particles, the mechanical performance of the composites is maintained. The stress–stretch curves are insensitive to the fraction of particles (Figure 4b and Figure S13). In particular, Young's modulus and strength remain nearly constant over the entire range of particle fractions, up to 80 wt% (Figure 4c,d).

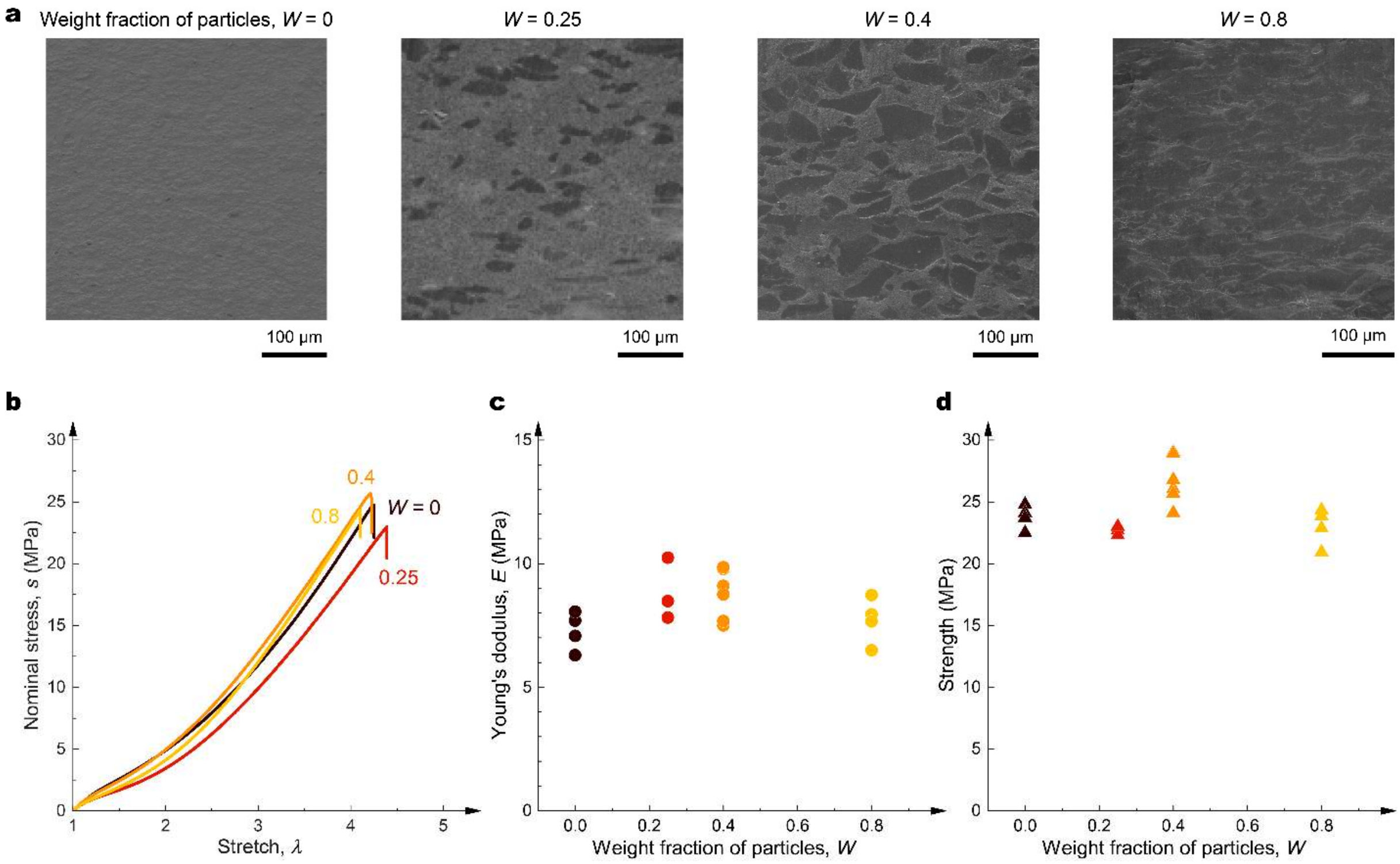

**Figure 4. A flaw-tolerant matrix recycles a high fraction of waste tire particles while maintaining strength. (a)** Scanning electron microscopy (SEM) images of the cross sections of composites with various weight fractions of particles. **(b)** Stress-stretch curves, **(c)** Young's modulus, and **(d)** strength as a function of the weight fraction of particles. For $W = 0$, 0.25, 0.4, and 0.8, $F = 0.25$, 0.2, 0.25, and 0.15, respectively. The initiator fraction is fixed as $C = 2\times10^{-3}$.

These results contrast sharply with previously reported composites containing waste tire particles, whose strength decreases markedly with increasing particle fraction (Figure 1d). As the fraction of particles increases, the number of flaw-like particles inevitably increases. These flaw-like particles may debond from the matrix or fracture internally and therefore initiate crack growth. Consequently, the strength of the previously reported composites decreases with increasing particle fraction. By contrast, in a flaw-tolerant matrix, flaw-like particles cannot initiate crack growth. So long as most particles remain adherent to the matrix and bear load, the composite maintains strength even if a small fraction of the particles act as flaws.

## 3. Discussion

Previous studies have applied flaw-tolerant rubbery materials to composites with stiff fibers[37] and to adhesion[38], but not to recycling. Recycling waste tire particles requires not only strong particle-matrix adhesion, but also a flaw-tolerant matrix. Such a matrix can incorporate large particles and high fractions of particles while maintaining properties. The flaw-tolerant matrix also enables composites to achieve both higher initiation toughness, $G_i$, and steady-state toughness, $G_c$, than a flaw-sensitive matrix (Figure S14). Moreover, a flaw-tolerant matrix reduces the need for every particle to be reliable, provided that the flaw-like particles are small compared to the fractocohesive length. The present work provides a flaw-tolerant principle that may aid efforts to recycle waste tires into new tires. Rather than relying on finer particles or stronger adhesion, flaw-tolerant recycling broadens the allowable particle size, thereby reducing the need to produce extremely fine particles.

The flaw tolerance of a matrix is characterized by its fractocohesive length. A material tolerates flaws smaller than its fractocohesive length, but not flaws larger than its fractocohesive length[32,33]. Here, the fractocohesive length is defined as $G_i/W_c$. We use $G_i$ because of a physical consideration: whether a matrix resists the initiation of crack growth depends on its initiation toughness, $G_i$, rather than the steady-state toughness, $G_c$. For a given material, the two values can differ greatly. For example, $G_c/G_i = 5.2$ for the flaw-tolerant matrix developed in this work (Figure S6), and $G_c/G_i = 10$ for the neat natural rubber with long chains and sparse crosslinks[39].

The design of a flaw-tolerant matrix itself deserves attention. Achieving a large fractocohesive length has gained increasing attention in the development of rubbery materials[40-42]. Previous studies on neat and inorganic particle-filled natural rubber have shown that a properly formed network with longer strands has a larger fractocohesive length[27,29,39]. For any given application, the fractocohesive length should be co-developed with other mechanical properties. For example, in this work, the matrix requires a large fractocohesive length while maintaining other properties, such as modulus, strength, stretchability, and crack resistance (Figure 2). Such codevelopment of fractocohesive length and other properties has not been a focus of past work. Ample room exists for mechanistic studies.

## 4. Conclusions

In summary, this work establishes the importance of flaw tolerance for recycling. Waste tire particles have degraded and variable properties[18-21]. Grinding waste tire rubber to extremely small particles is energy-intensive and costly[5,7,16,17]. Some particles inevitably act as flaws. Rather than requiring every particle to be reliable and small, flaw-tolerant recycling relies on a matrix designed so that flaw-like particles do not initiate macroscopic fracture. Such a matrix can incorporate large particles and high fractions of particles while maintaining properties. The flaw-tolerant principle may also be generalized to the recycling of other waste materials, such as plastics, elastomers, and composites. A flaw-tolerant matrix may also reduce the need to sort waste materials. For a given waste material, different chemistries and processing routes may be explored. We hope that the flaw-tolerant principle will aid efforts to recycle waste materials into high-performance products.

# 5. Experimental Section

## 5.1 Materials

Natural rubber latex (60 wt% rubber content) is purchased from Chemionics Corporation (OH, USA). The polyisoprene chains in latex particles have a high molecular weight of ~1,000,000 g/mol[43,44]. Dicumyl peroxide (DP, 329541) and toluene (244511) are purchased from Sigma-Aldrich. Carbon black (VULCAN® 3) is purchased from Cabot Corporation. Isopropanol (035913-4L) is purchased from Oakwood Products, Inc. All materials are used as received.

Waste tire particles with average sizes of 74 µm (PolyDyne 200) and 180 µm (PolyDyne 80) are purchased from Lehigh Technologies (Figure S8a,b). Waste tire particles ranging in size from 150 µm to 850 µm (20 Mesh Minus) are purchased from Tires to Green Recycling, LLC. The particles from Tires to Green Recycling, LLC are then sieved into different average sizes of 275 µm, 400 µm, and 750 µm, as confirmed by microscope images (Celestron, 5 MP Handheld Digital Microscope Pro) (Figure S8c-e). Waste tire particles consist primarily of natural rubber, styrene-butadiene rubber, polybutadiene, carbon black, and residual additives (Table S2). The density of waste tire particles is approximately $\rho_{tire}$ = 1.14 g cm$^{-3}$. All waste tire particles are washed with isopropanol three times and then air-dried before use.

## 5.2 Preparation of the flaw-tolerant matrix

The flaw-tolerant matrix is prepared using a wet-mixing process in which the long rubber chains are preserved (Figure S3). DP is dissolved in toluene. Then, the DP-toluene solution is mixed with natural rubber latex at a mass ratio of 6:1. The DP-toluene-water-latex mixture is stirred at 400 rpm overnight at room temperature. During mixing, latex particles dissolve into polyisoprene chains.

Separately, powders of carbon aggregates are dispersed in toluene at a mass ratio of 1:15. The dispersion is sonicated using a probe sonicator (YP-S500, Powdere; 500 W, 6 mm probe) at 25% amplitude. Sonication is performed in a pulsed mode (2 s on and 2 s off) for a total processing time of 40 min. The dispersion is maintained in an ice-water bath during sonication to minimize heating and toluene evaporation. During sonication, carbon agglomerates separate into aggregates, but the aggregates do not break.

The carbon-toluene dispersion is then added to the DP-toluene-water-latex mixture, which is further stirred with an overhead stirrer (ONiLAB OS20-S) equipped with a dissolver blade at 1000 rpm overnight at room temperature. The resulting mixture is poured into a Teflon petri dish. The mixture is kept in a fume hood at room temperature for five days to evaporate both toluene and water. As evaporation proceeds, rubber chains entangle, fill the gaps between nearby carbon aggregates, and occlude the open structure within each carbon aggregate. The mass of a sample is monitored during drying and reaches a plateau after four days. The dried, uncured compound is shaped at 95 °C for 30 minutes and then cured at 140 °C for 3 hours.

In this study, the volume fraction of carbon aggregates in the matrix is defined as $F = (m_{carbon}/\rho_{carbon})/(m_{carbon}/\rho_{carbon}+m_r/\rho_r)$, where $m_{carbon}$ and $\rho_{carbon}$ = 1.8 g cm$^{-3}$ are the mass and density of carbon aggregates, and $m_r$ and $\rho_r$ = 0.96 g cm$^{-3}$ are the mass and density of the solid natural rubber. The initiator fraction, $C$, is defined as the molar ratio of DP to the repeat units per polyisoprene chain. For the flaw-tolerant matrix, we choose $C = 2\times10^{-3}$ and vary $F$ from 0.15 to 0.25.

## 5.3 Preparation of the flaw-sensitive matrix

The compound of the flaw-sensitive matrix is prepared using the same processes of wet-mixing and drying as the flaw-tolerant matrix. Next, the compound is passed 100 times through an electric two-roll mill (130 mm Flat ULTRA Series X-DRIVE, Pepetools) with a roll gap of 1 mm, followed by shaping at 95 ºC for 30 minutes and curing at 140°C for 3 hours (Figure S4). Roll milling cuts the natural rubber chains short, and the resulting dangling ends of the short chains do not bear load. Therefore, dense crosslinks, $C = 10^{-2}$, are required. Meanwhile, we vary $F$ from 0.15 to 0.25.

### 5.4 Preparation of composites with a flaw-tolerant matrix

The composite with a flaw-tolerant matrix is prepared using a wet-mixing process (Figure S9). Waste tire particles are dispersed in toluene at a mass ratio of 1:8 using a stirrer. Separately, the carbon-toluene dispersion and the DP-toluene-water-latex mixture are prepared using the same procedures as those used for the flaw-tolerant matrix. Then, the particle-toluene dispersion is mixed with the carbon-toluene dispersion using an overhead stirrer for 30 minutes. The particle-carbon-toluene dispersion is then mixed with the DP-toluene-water-latex mixture using an overhead stirrer overnight.

The mixture is kept in a fume hood at room temperature for five days to evaporate toluene and water. As evaporation proceeds, rubber chains entangle, fill the gaps between nearby carbon aggregates, and occlude the open structure within each carbon aggregate. Meanwhile, rubber chains may diffuse into particles. By contrast, carbon aggregates, with sizes of 100-200 nm[45,46], are unlikely to diffuse into the particles. The compound is then shaped at 95 ºC for 30 minutes and cured at 140 ºC for 3 hours by hot-pressing.

In the composite, the weight fraction of particles is defined as $W = m_{\text{tire}}/(m_{\text{tire}}+m_{\text{carbon}}+m_{\text{r}})$, where $m_{\text{tire}}$ is the mass of particles. The volume fraction of carbon aggregates, $F$, is defined relative to the matrix, excluding waste tire particles. The initiator fraction, $C$, is defined in the same way as for the neat matrix.

### 5.5 Preparation of composites with a flaw-sensitive matrix

The composite with a flaw-sensitive matrix is prepared as follows (Figure S10). The natural rubber latex is first dried and then passed 100 times through the two-roll mill with a roll gap of 1 mm. Next, the milled rubber is dissolved in toluene at a mass ratio of 1:10 overnight using a stirrer, after which DP is added. Separately, the particle-carbon-toluene dispersion is prepared as described above. Then, the rubber-toluene solution is mixed with the particle-carbon-toluene dispersion using an overhead stirrer overnight. The mixture is kept in a fume hood at room temperature for five days to evaporate toluene and water. The compound is then shaped at 95 ºC for 30 minutes and cured at 140 ºC for 3 hours by hot-pressing.

### 5.6 Scanning electron microscopy (SEM)

SEM images are taken using a Zeiss Gemini 360 FE-SEM. To prevent charging, samples are coated with 10 nm-thick Pt/Pd (80/20 composition) using a metal sputter coater (EMS 150T S, Quorum Technologies). All SEM images are taken from the cross sections of the samples.

### 5.7 Stress-stretch curves

Stress-stretch curves are measured by uniaxial tension using dogbone-shaped samples. The dimensions of the gauge section of the sample are denoted by $L \times w \times t$, where $L$, $w$, and $t$ are the length, width, and thickness, respectively. In this study, $L = 16$ mm, $w = 2$ mm, and $t$ ranges from approximately 0.8 mm to 1.6 mm. The stretch is defined as $\lambda = l/L$, where $l$ is the deformed length of the sample. The nominal stress is defined as $s = P/(w \times t)$, where $P$ is the applied force. All samples are tested at a stretch rate of 1%/s until rupture. We calculate the modulus from a linear fit to the stress-stretch curve at small stretch ($\lambda <$ 1.02). The nominal stress and stretch at rupture define the strength, $S_f$, and stretchability, $\lambda_f$, respectively. The area under the stress-stretch curve up to rupture defines the work of fracture, $W_c$.

### 5.8 Pure shear tests

For each material, we prepare two samples: one with a notch and one without a notch (Figures S5 and S6). Both samples have the same width $w$, height $H$, and thickness $t$. In this study, $w = 50$ mm, $H = 10$ mm, and $t$ ranges from approximately 0.8 mm to 1.6 mm. For the notched sample, a 15-mm-long precrack is introduced. The stretch is defined as $\lambda = h/H$, where $h$ is the deformed height of the sample. The nominal stress is defined as $s = P/(w \times t)$. For the notched sample, the crack growth process is recorded as a video. From the video, we determine the stretch at the initiation of crack growth, $\lambda_i$. From the stress-stretch curve, we determine the stretch at the maximum nominal stress, $\lambda_c$. Then, the strain energy per unit volume, $W(\lambda_i)$ and $W(\lambda_c)$, are calculated by integrating the stress-stretch curve of the unnotched sample from $\lambda = 1$ to $\lambda = \lambda_i$ and from $\lambda = 1$ to $\lambda = \lambda_c$, respectively. The initiation toughness and steady-state toughness are calculated as $G_i = H \times W(\lambda_i)$ and $G_c = H \times W(\lambda_c)$, respectively. All tests are performed at a stretch rate of 1%/s. The fractocohesive length is defined as $G_i/W_c$.

### 5.9 AI-assisted illustration

The initial illustration for Figure 1a was generated using ChatGPT (OpenAI) and subsequently modified by the authors. The final illustration was reviewed and verified by the authors.

## Acknowledgements

This research is supported by the National Science Foundation through the Harvard University Materials Research Science and Engineering Center DMR-2011754.

## Author contributions

C.L. conceived the project. C.L., Y.Z., Z.C., Y.K. and Z.S. designed the study and analyzed the results. C.L. conducted experiments and data analysis. Y.Z. helped with the preparation of the flaw-sensitive matrix and composites. J.M. and M.W.M.T. took the SEM images. M.Y. helped with the preparation of the flaw-tolerant matrix in the initial stage. Z.S. and Y.K. supervised the research. C.L., Y.Z., Z.C., Y.K. and Z.S. wrote the manuscript with input from all authors. All authors read and approved the final article.

## Conflicts of Interest

The authors declare the following competing interest: A provisional patent application related to this work has been filed by Harvard University. The authors declare no other competing interests.

## Data availability

The data that support the findings of this study are available in the Supporting Information of this article.

## References

1 Abbas-Abadi, M. S., Kusenberg, M., Shirazi, H. M., Goshayeshi, B. & Van Geem, K. M. Towards full recyclability of end-of-life tires: Challenges and opportunities. *J. Clean. Prod.* **374**, 134036 (2022).

2 Dabic-Miletic, S., Simic, V. & Karagoz, S. End-of-life tire management: a critical review. *Environ. Sci. Pollut. Res.* **28**, 68053-68070 (2021).

3 Grammelis, P., Margaritis, N., Dallas, P., Rakopoulos, D. & Mavrias, G. A review on management of end of life tires (ELTs) and alternative uses of textile fibers. *Energies* **14**, 571 (2021).

4 Shu, X. & Huang, B. Recycling of waste tire rubber in asphalt and portland cement concrete: An overview. *Constr. Build. Mater.* **67**, 217-224 (2014).
5 Formela, K. Sustainable development of waste tires recycling technologies–recent advances, challenges and future trends. *Adv. Ind. Eng. Polym. Res.* **4**, 209-222 (2021).
6 Sathiskumar, C. & Karthikeyan, S. Recycling of waste tires and its energy storage application of by-products–a review. *Sustain. Mater. Technol.* **22**, e00125 (2019).
7 Fazli, A. & Rodrigue, D. Recycling waste tires into ground tire rubber (GTR)/rubber compounds: A review. *J. Compos. Sci.* **4**, 103 (2020).
8 Georgiou, D., Sun, D., Liu, X. & Athanasiou, C. E. Suppressing mechanical property variability in recycled plastics via bioinspired design. *Proc. Natl Acad. Sci. USA* **122**, e2502613122 (2025).
9 Archibong, F. N., Sanusi, O. M., Médéric, P. & Aït Hocine, N. An overview on the recycling of waste ground tyre rubbers in thermoplastic matrices: Effect of added fillers. *Resour. Conserv. Recycl.* **175**, 105894 (2021).
10 Nuzaimah, M., Sapuan, S., Nadlene, R. & Jawaid, M. Recycling of waste rubber as fillers: A review. *IOP Conf. Ser.: Mater. Sci. Eng.* **368**, 012016 (2018).
11 Formela, K. Waste tire rubber-based materials: Processing, performance properties and development strategies. *Adv. Ind. Eng. Polym. Res.* **5**, 234-247 (2022).
12 Ramarad, S., Khalid, M., Ratnam, C. T., Chuah, A. L. & Rashmi, W. Waste tire rubber in polymer blends: A review on the evolution, properties and future. *Prog. Mater. Sci.* **72**, 100-140 (2015).
13 Bianco, I., Shirzadi, R. & Blengini, G. A. Closed-loop recycling of End-of-Life tires via devulcanized rubber: Life Cycle Assessment and circularity evaluation with durability trade-offs. *Journal of Cleaner Production* **571**, 148760 (2026).
14 Wu, Q., Zhang, Q., Chen, X., Song, G. & Xiao, J. Life cycle assessment of waste tire recycling: Upgraded pyrolytic products for new tire production. *Sustainable Production and Consumption* **46**, 294-309 (2024).
15 Valentini, F. & Pegoretti, A. End-of-life options of tyres. A review. *Advanced Industrial and Engineering Polymer Research* **5**, 203-213 (2022).
16 Adhikari, B., De, D. & Maiti, S. Reclamation and recycling of waste rubber. *Prog. Polym. Sci.* **25**, 909-948 (2000).
17 Karger-Kocsis, J., Mészáros, L. & Bárány, T. Ground tyre rubber (GTR) in thermoplastics, thermosets, and rubbers. *J. Mater. Sci.* **48**, 1-38 (2013).
18 Baldwin, J. M. & Bauer, D. R. Rubber oxidation and tire aging-a review. *Rubber Chem. Technol.* **81**, 338-358 (2008).
19 Baldwin, J. M., Bauer, D. R. & Ellwood, K. R. Rubber aging in tires. Part 1: Field results. *Polym. Degrad. Stab.* **92**, 103-109 (2007).
20 Ahagon, A., Kida, M. & Kaidou, H. Aging of tire parts during service. I. Types of aging in heavy-duty tires. *Rubber Chem. Technol.* **63**, 683-697 (1990).
21 Stabile, P., Mastinu, G., Gobbi, M. & Bardini, P. Tire aging: A state-of-the-art review. *Eng. Sci. Technol. Int. J.* **69**, 102101 (2025).
22 Wang, X., Du, J., Zhang, X. & Jia, D. Mechano-chemical modification of waste rubber powder with 2-mercatobenzothiozole and 3-aminopropyltriethoxysilane. *Macromol. Res.* **26**, 403-409 (2018).
23 Guo, L. *et al.* Enhancing interface adhesion between waterjet-produced rubber powders and natural rubber by wet-mixing method. *J. Appl. Polym. Sci.* **141**, e55555 (2024).
24 Sun, Y., Yan, X., Liang, H., Böhm, G. & Jia, L. Rubber recycling: mending the interface between ground rubber particles and virgin rubber. *ACS Appl. Mater. Interfaces* **12**, 47957-47965 (2020).
25 Kong, L. *et al.* High performance, self-adhesive and recyclable dynamic crosslinked waste rubber particle blends toward upcycling of waste rubber. *Green Chem.* **26**, 1523-1532 (2024).
26 Wiśniewska, P., Wang, S. & Formela, K. Waste tire rubber devulcanization technologies: State-of-the-art, limitations and future perspectives. *Waste Manage.* **150**, 174-184 (2022).

27 Tan, M. W. M. *et al.* Amplifying toughness in silica-reinforced natural rubber by preserving long chains. *Proc. Natl Acad. Sci. USA* **123**, e2530834123 (2026).
28 Tan, M. W. M. *et al.* Composite of long polymer chains and structured silica particles achieves high modulus and low hysteresis. *ACS Mater. Au* **6**, 604-618 (2026).
29 Li, C. *et al.* Resolving toughness-modulus conflict in carbon black reinforced natural rubber by preserving long chains. *Soft Matter* **22**, 518-531 (2026).
30 Bao, X. *et al.* Low-intensity mixing process of high molecular weight polymer chains leads to elastomers of long network strands and high fatigue threshold. *Soft Matter* **19**, 5956-5966 (2023).
31 Rubinstein, M. & Colby, R. H. *Polymer physics*. Vol. 23 (Oxford University Press, 2003).
32 Yang, C., Yin, T. & Suo, Z. Polyacrylamide hydrogels. I. Network imperfection. *J. Mech. Phys. Solids* **131**, 43-55 (2019).
33 Chen, C., Wang, Z. & Suo, Z. Flaw sensitivity of highly stretchable materials. *Extreme Mech. Lett.* **10**, 50-57 (2017).
34 Torquato, S. *Random heterogeneous materials: microstructure and macroscopic properties*. Vol. 16 (Springer, 2002).
35 Karasek, L., Meissner, B., Asai, S. & Sumita, M. Percolation concept: polymer-filler gel formation, electrical conductivity and dynamic electrical properties of carbon-black-filled rubbers. *Polym. J.* **28**, 121-126 (1996).
36 Rintoul, M. D. & Torquato, S. Precise determination of the critical threshold and exponents in a three-dimensional continuum percolation model. *J. Phys. A: Math. Gen.* **30**, L585-L592 (1997).
37 Chen, X., Liu, F., Lu, Y., Suo, Z. & Tang, J. Stretchable materials show high crack-initiation resistance that exceeds soft tissues. *Mater. Today* **93**, 103199 (2026).
38 Chen, B., Yang, J., Bai, R. & Suo, Z. Molecular staples for tough and stretchable adhesion in integrated soft materials. *Adv. Healthc. Mater.* **8**, 1900810 (2019).
39 Nian, G. *et al.* Natural rubber with high resistance to crack growth. *Nat. Sustain.* **8**, 692-701 (2025).
40 Hartquist, C. M. *et al.* Extraordinarily high fractocohesive lengths in polymer-like networks. *Math. Mech. Solids* **0**, 0 (2026).
41 Sun, F. *et al.* A versatile microporous design toward toughened yet softened self-healing materials. *Adv. Mater.* **36**, 2410650 (2024).
42 Ma, J. *et al.* Designing ultratough single-network hydrogels with centimeter-scale fractocohesive lengths via inelastic crack blunting. *Adv. Mater.* **36**, 2311795 (2024).
43 Hayeemasae, N., Soontaranon, S. & Masa, A. Structure–property relationships of different natural rubber grades. *Prog. Rubber Plast. Recycl. Technol.* **41**, 36-52 (2025).
44 Westall, B. The molecular weight distribution of natural rubber latex. *Polymer* **9**, 243-248 (1968).
45 Vilgis, T. A. Time scales in the reinforcement of elastomers. *Polymer* **46**, 4223-4229 (2005).
46 Blackford, D. B. & Simons, G. R. Particle size analysis of carbon black. *Part. Charact.* **4**, 112-117 (1987).